\documentstyle[preprint,version2,aps]{revtex}

\begin{document}
\draft
\begin{title}
%%%%%%%%%%%%%%%%%%%%%%%%%%%%%%%%%%%%%%%%%%%%%%%%%%%%%
Renormalized Harmonic-Oscillator Description of \\
Confined Electron Systems with Inverse-Square Interaction
%%%%%%%%%%%%%%%%%%%%%%%%%%%%%%%%%%%%%%%%%%%%%%%%%%%%%
\end{title}
\author{Norio Kawakami}
\begin{instit}
Yukawa Institute for Theoretical Physics,
Kyoto University, Kyoto 606, Japan
\end{instit}
%\moreauthors{}
%\receipt{ 1993}
\begin{abstract}
%%%%%%%%%%%%%%%%%%%%%%%%%%%%%%%%%%%%%%%%%%%%%%%%%%%%%%%%%%%%%%
An integrable model for SU($\nu$) electrons with
inverse-square interaction is studied for the
system with  confining harmonic potential.  We develop a
new description of the spectrum
based on the {\it renormalized harmonic-oscillators}
which incorporate interaction effects via the repulsion
of energy levels. This approach enables a systematic
treatment of the excitation spectrum as well as the ground-state
quantities.
%%%%%%%%%%%%%%%%%%%%%%%%%%%%%%%%%%%%%%%%%%%%%%%%%%%%%%%%%%%
\end{abstract}
\vskip 20mm

\vspace{1 cm}
% for preprint style only
%\pacs{PACS numbers: ******}
\newpage
\narrowtext
%%%%%%%%%%%%%%%%%%%%%%%%%%%
Quantum integrable systems with inverse-square ($1/x^2$)
interaction have been studied extensively with renewed interest
\cite{calo,sutha,hala,shas,kura,ky,kawa,kiwata,ha,hikami,poly}.
Though various models including multicomponent cases have
been studied in detail for periodic boundary conditions,
the energy spectrum for confined systems has not been
investigated systematically so far. In fact various  investigations
concerning confined systems have been mainly devoted to
the single-component case \cite{calo,sutha}.
Besides interest in the integrability there have
been some attempts to apply the confined models to
conductance-oscillation phenomena in  mesoscopic
systems \cite{tewari,jp,vacek}, in which the exact ground state
has been found for the {\it electron} model  \cite{vacek}.
It is  desirable to develop  systematic treatments of
the energy spectrum for the class
of the integrable  multicomponent models with confining potential.

In this letter we wish to propose
a systematic description of the energy spectrum
for the class of one-dimensional multicomponent inverse-square
systems with harmonic confinement.  The essence of the idea is
that the energy spectrum is expressed in terms of
the {\it renormalized  harmonic oscillators}  (RHO)
for which the interaction effects are
taken into account by the renormalized quantum numbers
for oscillators.

To illustrate the idea, let us start with the
integrable spinless-fermion model with
confining harmonic potential \cite{calo,sutha},
%%%%%%%%%%%%%%%%%%%%%%%%%%%%%%%%%%%%%
\begin{equation}
H = - \frac{1}{2} \sum_{i=1}^{N} \frac{\partial^2}{\partial x_i^2}
    + \frac{1}{2} \sum_{i=1}^{N} \omega_0^2 x_i^2
    + \sum_{j>i} \frac{\lambda(\lambda+1)}{(x_j-x_i)^2},
\end{equation}
%%%%%%%%%%%%%%%%%%%%%%%%%%%%%%%%%%%%%%%
with $N$ being the total electron number,
where the dimensionless coupling is assumed to be $\lambda> 0$.
Let us first recall the key feature common to
inverse-square systems: the interaction effect
in (1) causes  the {\it repulsion}
of energy levels, which uniformly enlarges the spacing of
adjacent quantum numbers  $\lambda+1$ times as large as
non-interacting case \cite{sutha}. In the periodic case, for example,
this repulsion gives rise to the step-wise phase shift in  the
scattering matrix, which can be treated systematically via the
asymptotic Bethe ansatz (BA) \cite{sutha,kawa}.
In  such cases the renormalized quantities coincide
with the rapidities (or quasi-momentums) in the BA.
Though the BA solution can not be applied directly to
the confined systems because of lack of translational invariance,
we wish to develop the systematic treatment
by taking into account the above key feature of the level
repulsion.

Similarly to the periodic case, the repulsion of energy
levels in the confined case is
expressed by the renormalization of the quantum
numbers for oscillators. In the  RHO
approach proposed here, it is conjectured that
the energy is expressed in the formula for free harmonic
oscillators ($\hbar$ is written explicitly here),
%%%%%%%%%%%%%%%%%%%%%%%%%%%%%%%%%%%%%%%%
\begin{equation}
E= \hbar \omega_0
\sum_{j=1}^{N} (n_j +{1\over 2}),
\end{equation}
%%%%%%%%%%%%%%%%%%%%%%%%%%%%%%%%%%%%%%%%%
and the effect of interaction is  incorporated
into the {\it renormalized  quantum number}
of oscillators, $n_j$, which is an analogue of the
rapidity in the periodic case.
This quantity is not necessarily an integer  once
the correlation is introduced, but changes continuously
according to the interaction strength.
Noting that the repulsion of  energy levels
are described by introducing the step function as
$\lambda \theta(n_i-n_j)$ \cite{sutha,kawa},
we find the quantum number $n_j$
to be determined by the BA-type equation,
%%%%%%%%%%%%%%%%%%%%%%%%%%%%%%%%%%%%%%%%%%%
\begin{equation}
n_j=  I_j^{(1)} + \lambda \sum_{l \neq j} \theta (n_j- n_l),
\end{equation}
%%%%%%%%%%%%%%%%%%%%%%%%%%%%%%%%%%%%%
where $I_j^{(1)}(= 0, 1, 2, \cdots$) is the bare
quantum number, and we define the step function as:
$\theta(x)=1$ for $x>0$ and   $\theta(x)=0$ for $x \leq 0$.
It is seen from (3) that the spacing of the renormalized
quantum numbers $n_j$ is indeed enlarged $\lambda+1$ times
as large as the non-interacting case, as should be expected.
{}From (2) and (3) the energy for an arbitrary excited state
turns out to take the simple form,
%%%%%%%%%%%%%%%%%%%%%%%%%%%%%%%%%%%%
\begin{equation}
E=  \hbar \omega_0 [ {1 \over  2} \lambda N(N-1) +
\sum_{j=1}^{N} (I_j^{(1)}+{1 \over 2})].
\end{equation}
%%%%%%%%%%%%%%%%%%%%%%%%%%%%%%%%%%%%%%%%%%%%%
The ground state is given by the set of quantum numbers
$I_j^{(1)}= 0, 1, \cdots, N-1$, resulting in the energy
$E_g=  \hbar \omega_0 [N^2/2 + \lambda N(N-1)/2]$.
Also, particle-hole excitations can be described by changing
the quantum numbers $I_j^{(1)}$ from those for the ground state.
It is seen from (4) that particle-hole excitations
 do not include any interaction effects.  The above
formula (4) reproduces the exact results \cite{calo,sutha}.

Let us now turn to the SU($\nu$) electron model with the
internal spin degrees of freedom. The Hamiltonian reads
\cite{ha,hikami,poly,vacek},
%%%%%%%%%%%%%%%%%%%%%%%%%%%%%%%%%%%%%
\begin{equation}
H = - \frac{1}{2} \sum_{i=1}^{N} \frac{\partial^2}{\partial x_i^2}
    + \frac{1}{2} \sum_{i=1}^{N} \omega_0^2 x_i^2
    + \sum_{j>i} \frac{\lambda(\lambda+P_{ij}^{\alpha\beta})}
{(x_j-x_i)^2},
\end{equation}
%%%%%%%%%%%%%%%%%%%%%%%%%%%%
where $P_{ij}^{\alpha\beta}$ is the spin-exchange operator of two
particles with spin indices $\alpha$
and $\beta$ ($=1,2, \cdots, \nu$). Although
the above model is known to be integrable \cite{hikami,poly},
the excitation spectrum has not been obtained so far while
the periodic version of the model  has been already studied
in detail \cite{ha}.  Note that the ground-state
wavefunction for the above model
includes the Vandermonde determinantal products,
the power of which is raised up to $\lambda+1$
for the particles with same spins and to $\lambda$
for different spins \cite{ha,vacek}.

The idea of RHO description turns out to be
still applicable to  the SU($\nu$) model.
Although ordinary techniques in  the nested-BA
are not straightforwardly applied,  we can use
the modified nested-BA  developed for
the $1/r^2$ systems, the detail of which is given in \cite{kawaa}.
As a consequence, we necessarily introduce the
set of renormalized quantum numbers $n_j^{(\alpha)}$
($\alpha=1,2, \cdots, \nu$) which are determined
consistently by the BA-like equations,
%%%%%%%%%%%%%%%%
\begin{equation}
 n_j^{(1)} =   I_j^{(1)}
- \sum_{m}^{M_2} \theta(n_j^{(1)} -n_m^{(2)})
+ \lambda \sum_{l}^{M_1}\theta(n_j^{(1)}-n_l^{(1)}),
\end{equation}
%%%%%%%%%%%%%%%%
\begin{equation}
\sum_{l}^{M_\alpha}\theta(n_m^{(\alpha)}-n_l^{(\alpha)}) +
I_{m}^{(\alpha)}
=  \sum_{j}^{M_{\alpha-1}} \theta(n_m^{(\alpha)}-n_j^{(\alpha-1)})
+\sum_s^{M_\alpha+1} \theta(n_m^{(\alpha)}-n_s^{(\alpha+1)}),
\end{equation}
%%%%%%%%%%%%%%%%%%%%
for $2 \leq \alpha \leq \nu$,
where non-negative integers $I_j^{(\alpha)}$
($=0, 1, \cdots$) are the bare quantum numbers
which classify $\nu$ species of elementary excitations.
Here we have introduced $M_\alpha=\sum_{\beta=\alpha}^\nu N_\beta$,
where $N_\beta$ is the number of electrons with $\beta$ spin
($M_1=N=\sum_{\beta=1}^{\nu} N_{\beta}$).
In the RHO approach the total energy is given in the
formula of harmonic oscillators as in (2) with $n_j=n_j^{(1)}$,
which is straightforwardly evaluated using (6) and (7) as,
%%%%%%%%%%%%%%%%%%%%%%%%%%%%%
\begin{equation}
E =  \hbar \omega_0 [{1 \over 2} \lambda N(N-1)
 + \sum_{\alpha=1}^{\nu}
({1 \over 2} N_\alpha^2 - {1 \over 2} M_\alpha(M_\alpha-1)
+ \sum_{j=1}^{M_\alpha} I_j^{(\alpha)})].
\end{equation}
%%%%%%%%%%%%%%%%%%%%%%%%%%%%%%%%%%
The ground state for SU($\nu$) singlet is given by  the set of
quantum numbers $I_j^{(\alpha)}= 0, 1, \cdots, M_\alpha-1$.
The corresponding ground-state energy is computed as
%%%%%%%%%%%%%%%%%%%%%%%
\begin{equation}
E_g=  \hbar \omega_0 [
{1 \over 2} \lambda N(N-1)
 + {1 \over 2} \sum_{\alpha=1}^{\nu} N_\alpha^2].
\end{equation}
%%%%%%%%%%%%%%%%%%%%%%%
which reproduces the exact results already known for the
SU(2) case \cite{vacek}.
Remarkably enough, all the interaction effects are
incorporated via the first term in (8), and
excitations with the fixed number of electrons do not
include any effects of interactions,
as is the case for the spinless-fermion case.
Consequently, if the number of electrons is {\it fixed},
the spectrum is indeed described by free oscillators with the
{\it bare} frequency $\hbar \omega_0$.
%%%%%%%%%%%%%%%%%%%%%%%
%To avoid confusions, let us stress that this does not mean
%the system to consist of free oscillators.
%%%%%%%%%%%%%%%%%%%%%%%%%%
This consequence for the confined system is quite contrasted to the
periodic model with inverse-square  interaction where
the frequency (or velocity) is strongly
 renormalized by the interaction \cite{ha}.

We now wish to classify the excitation spectrum in more
convenient form to see symmetry property.
Let $m_\alpha$ be the change of  the
electron number with $\alpha$-th spin
($\Delta N= \sum_{\alpha=1}^{\nu} m_\alpha$). The excitation
spectrum then reduces to
%%%%%%%%%%%%%%%%%%%%%%%%%%%%%%%%
\begin{equation}
\Delta E = \hbar \omega_0 [{1 \over 2} \,
\vec m^t {\bf \tilde C} \vec m + \sum_{\alpha=1}^\nu l^{(\alpha)}]
\end{equation}
%%%%%%%%%%%%%%%%%%%%%%%%%%%%%%%
with the $\nu\times \nu$ matrix
\begin{equation}
{\bf \tilde C}=
\left(
\matrix {\lambda+1  & \lambda      & \cdots  & \lambda   \cr
         \lambda   &  \lambda+1      & \ddots    & \vdots   \cr
        \vdots       & \ddots  & \ddots & \lambda  \cr
        \lambda       & \cdots   &  \lambda    & \lambda+1       \cr}
\right),
\end{equation}
%%%%%%%%%%%%%%%%%%%%%%%%
%\lambda +\delta_{ij},
%\end{equation}
%%%%%%%%%%%%%%%%%%%%%%%%%%%%%
where we have dropped the chemical-potential term from the
expression. Here the integer-valued vector is
$ \vec m=(m_1, \cdots, m_\nu)$, and non-negative
integers $l^{(\alpha)}$ label excitations of particle-hole type.
The above excitation spectrum has been  expressed in terms
of quantum numbers $m_\alpha$ for electrons.
We note that there is an alternative expression
based on quantum numbers $M_j$ ($j=1,3, \cdots, \nu$)
concerning  spin and charge excitations.
Quantum numbers in these two bases
are mutually transformed by the linear transformation such as
$\vec M ={\bf U}^{-1} \vec m$ with the $\nu \times \nu$
matrix $U_{\alpha\beta}=\delta_{\alpha\beta}
-\delta_{\alpha(\beta-1)}$ \cite{kawaa}.  By this transformation,
the matrix $\tilde {\bf C}$ is converted into the matrix
in the basis of spin-charge excitations,
%%%%%%%%%%%%%%%%%%%%%%%%%%%%%%%%
\begin{equation}
{\bf  C}=
\left(
\matrix {\lambda+1  & -1      &  &    \cr
             -1    &  2      & \ddots    &     \cr
                    & \ddots  & \ddots   &  -1  \cr
                    &         &  -1      & 2    \cr}
\right).
\end{equation}
%%%%%%%%%%%%%%%%%%%%%%%%%%%%%%%%%%%%
The above expression clearly demonstrates
that the excitation spectrum of the present system
has the same form as for {\it chiral} SU($\nu$) electron systems.
In fact the above matrix ${\bf C}$ has been previously used
to classify the low-energy spectrum for
translationally invariant electron systems of the length
$L$ and with the velocity $v$  ($\hbar \omega_0$
is replaced by $2 \pi v /L$ in (10)) \cite{kawaa}.
Therefore  the present model with confining potential
naturally reproduces the spectrum for the SU($\nu$) {\it chiral}
Luttinger liquid in the limit of
$\omega_0\rightarrow 0$, the critical
behavior of which is described by holomorphic piece of
U(1) Kac-Moody algebra (charge excitation)
and SU($\nu$) Kac-Moody algebra (spin excitation) \cite{bpz,review}.

In summary we have developed the renormalized
harmonic-oscillator description of the
energy spectrum for the integrable SU($\nu$) model
with inverse-square interaction.
This approach has enabled a systematic treatment of the
excitation spectrum.  Renormalized quantum numbers in the present
formalism are conserved quantities, which
are analogues of rapidities in the ordinary BA solution.
So, these key quantities should be related to
conserved charges which ensure the integrability of the model.
It is interesting to find the explicit relationship between
 renormalized quantum numbers and conserved charges.
Also, it remains open to establish the microscopic foundation
of the present approach by  constructing
eigenfunctions explicitly.

Fruitful discussions with
H. Fukuyama,  Y. Kuramoto,  A. Okiji, J. Solyom, K. Vacek
and  S.-K. Yang are  gratefully acknowledged.
This work is partly supported by Grant-in-Aid from
the Ministry of Education, Science and Culture
and also by Monbusho International Scientific Research
Program.

%%%%%%%%%%%%%%%%%%%%%%%%
\newpage
%%%%%
%%%%%%%%%%%%%%%%%%%%%%%%%%%%%%%%%%%%%%%%%%%%%%%%%

\newpage
%%%%%%%%%%%%%%%%%%%%%%%%%%%%%%%%%%%%%%%%
\end{document}